\newcommand{\be}{\begin{equation}}
\newcommand{\ee}{\end{equation}}
\newcommand{\bea}{\begin{eqnarray}}
\newcommand{\eea}{\end{eqnarray}}
\newcommand{\bean}{\begin{eqnarray*}}

\newcommand{\eean}{\end{eqnarray*}}
\font\upright=cmu10 scaled\magstep1
\font\sans=cmss10
\newcommand{\ssf}{\sans}
\newcommand{\stroke}{\vrule height8pt width0.4pt depth-0.1pt}
\newcommand{\Z}{\hbox{\upright\rlap{\ssf Z}\kern 2.7pt {\ssf Z}}}

\newcommand{\C}{{\rlap{\rlap{C}\kern 3.8pt\stroke}\phantom{C}}}
\newcommand{\R}{\hbox{\upright\rlap{I}\kern 1.7pt R}}
\newcommand{\CP}{\C{\upright\rlap{I}\kern 1.5pt P}}
\newcommand{\PP}{\hbox{\upright\rlap{I}\kern 1.5pt P}}

\newcommand{\identity}{{\upright\rlap{1}\kern 2.0pt 1}}

\newcommand{\HH}{\mbox{\hbox{\upright\rlap{I}\kern 1.7pt H}}}

\newcommand{\fr}{\frac}
\newcommand{\lm}{\lambda}
\newcommand{\ra}{\rightarrow}

\newcommand{\pr}{\partial}
\newcommand{\hs}{\hspace{5mm}}

\newcommand{\acc}{\\[3mm]}

\input{epsf}

\documentstyle[12pt,a4wide]{article}

\begin{document}
\title{\vskip -90pt
\begin{flushright}
\end{flushright}\vskip 30pt
{\bf \large \bf Bogomolny Yang-Mills-Higgs Solutions  in (2+1) anti-de 
Sitter Space}
\author{Theodora Ioannidou\\[10pt]  
\\{\normalsize  {\sl Institute of Mathematics, University of Kent at
 Canterbury,}}\\
{\normalsize {\sl Canterbury, CT2 7NF, U.K.}}\\
{\normalsize{\sl Email : T.Ioannidou@ukc.ac.uk}}\\}}
\maketitle

\begin{abstract}
This paper investigates an integrable system which is related to hyperbolic
monopoles; ie the Bogomolny Yang-Mills-Higgs equations in (2+1) anti-de Sitter 
space which are integrable and whose solutions can be obtained using 
analytical methods.
In particular, families of soliton solutions have been constructed explicitly
and their dynamics has been investigated in some detail.\\
\end{abstract}

\renewcommand{\thefootnote}{\arabic{footnote}}
\setcounter{footnote}{0}

{\large{\bf I. Introduction}}\\

Static BPS monopoles are solutions of a nonlinear elliptic partial
 differential equation on some three-dimensional Riemannian 
manifold.
Most work on monopoles has dealt with the case when this manifold is
Euclidean space $\R^3$ since the equations are integrable and geometrical
techniques can be applied.
[The introduction of  time dependence destroys the integrability].
In addition, the monopole equations on hyperbolic space $\HH^3$ are also 
integrable \cite{At} and often hyperbolic monopoles turn out to be easier 
to study than the Euclidean (see, for example, \cite{IS}).
Moreover, recently, it has been rigorously established \cite{JN} that
in the limit as the curvature of hyperbolic space tends to
 zero then Euclidean monopoles are recovered.
In this paper, we consider  an integrable system \cite{W1} which is related to 
hyperbolic monopoles and follows from replacing the positive definite
 space $\HH^3$ by a Lorentzian version, ie the anti-de Sitter space.
In recent years, the $n$-dimensional anti-de Sitter spacetime has been
of continuing interest since it is the base of M-theory and a source 
of simple examples studying methods and spacetime concepts both on
classical and quantum level.
It also arises as the natural ground state of gauged supergravity
 theories when quantized \cite{Gary}.

The Bogomolny version of Yang-Mills-Higgs equations for 
Yang-Mills-Higgs fields on a three-dimensional 
Riemannian manifold (${\cal M}$) with gauge group $SU(2)$ have the form
\be
D_i \Phi=\fr{1}{2\sqrt{|g|}} g_{ij} \,\epsilon^{jkl} F_{kl}.
\label{Bog}
\ee
Here $A_k$, for $k=0,1,2$, is the $su(2)$-valued gauge potential, 
with field strength $F_{ij}=\pr_iA_j-\pr_jA_i+[A_i,A_j]$ and  
$\Phi=\Phi(x^\mu)$ is the $su(2)$-valued Higgs field; while
 $x^\mu=(x^0,x^1,x^2)$ represent the local coordinates on $M$.
The action of the covariant derivative $D_i=\pr_i+A_i$ on $\Phi$ is:
 $D_i\Phi=\pr_i\Phi+[A_i,\Phi]$.
Equation (\ref{Bog}) is integrable in the sense that a Lax pair exists 
for constant curvature.
In particular, the solutions of (\ref{Bog}) correspond to Euclidean 
or hyperbolic BPS monopoles when $({\cal M},g)$ is Euclidean $\R^3$ or 
hyperbolic $\HH^3$  space. 

There are two curved spacetimes with constant curvature: (i) the de Sitter 
space with positive scalar curvature and (ii) the anti-de Sitter space
with negative curvature.
By definition the (2+1)-dimensional anti-de Sitter space is the universal
covering space of the hyperboloid ${\cal H}$ satisfied by the equation
\be
U^2+V^2-X^2-Y^2=1
\ee
with metric given by
\be
ds^2=-dU^2-dV^2+dX^2+dY^2.
\ee
By parametrizing the hyperboloid ${\cal H}$ by
\bea
U&=&\sec\rho\cos\theta\nonumber\\
V&=&\sec\rho\sin\theta\nonumber\\
X&=&\tan\rho\cos\phi\nonumber\\
Y&=&\tan\rho\sin\phi
\eea
for $\rho \in [0,\pi/2)$, the corresponding metric takes the form
\be
ds^2=\sec^2\rho \left(-d\theta^2+d\rho^2+\sin^2\rho\, d\phi^2\right).
\ee
The spacetime contains closed timelike curves, due to the periodicity
of $\theta$ (for more details, see Ref. \cite{HE}).
In fact, anti-de Sitter space (as a manifold) is the product of an
 open spatial disc with $\theta$ and constant curvature equal 
to minus six; where $(\rho,\phi)$ correspond to polar coordinates
and $\theta \in R$ being the time.
Null spacelike infinity ${\cal I}$ consists of the timelike 
cylinder $\rho=\pi/2$ and this surface is never reached by 
timelike geodesics.

\begin{figure}
\vskip 0.5cm
\centerline{
\put(90,97){$\theta$}
\put(80,50){$r=0$}
\put(60,60){$t$}
\put(85,3){$\rho$}
\put(7,100){$(\rho,\phi,\theta)$}
\put(7,60){$(r,x,t)$}
\epsfxsize=3cm\epsfysize=3cm\epsffile{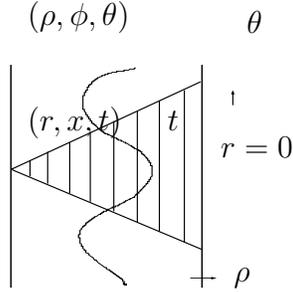}}
\vskip 0.5cm
\caption{The Penrose diagram of anti-de Sitter space.
The boundary of anti-de Sitter is the boundary of the cylinder.}
\label{fig-ads}
\end{figure}

If the Poincar\'e coordinates $(r,x,t)$ for $r>0$ are defined as
\bea
r&=&\fr{1}{U+X}\nonumber\\
x&=&\fr{Y}{U+X}\nonumber\\
t&=&\fr{-V}{U+X}
\label{twis}
\eea
the metric simplifies to the following form
\be
ds^2=r^{-2}(-dt^2+dr^2+dx^2).
\label{Pmet}
\ee
Note that, the Poincar\'e coordinates cover a small part of anti-de Sitter 
space, ie that corresponding to half of the hyperboloid ${\cal H}$  
for $U+X>0$; which is the shaded region in FIG. \ref{fig-ads}.
The surface $r=0$ is part of infinity ${\cal I}$.

Hitchin \cite{Hitchin} show that the minitwistor space corresponding to 
Poincar\'e space (\ref{Pmet}) is $CP^1\times CP^1$ and can be visualized
as a quadric Q in $CP^3$; while the points of spacetime correspond 
to certain plane sections (conics) of $Q$ with space $CP^3$.
The relevant conics which have to be real and nondegenerate, are given 
by the expression \cite{W1}
\be
\omega=v-r^2\,(\mu-u)^{-1}
\label{omega}
\ee
where $(\omega,\mu)$ are standard coordinates on the two $CP^1$ factor
of $Q$, while $u=x+t$ and $v=x-t$.
Note that the Poincar\'e coordinates $(r,x,t)$ cover all of the 
space of these conics (which is the top half of $RP^3$) expect for
a set of measure zero.
In order to see the correspondence between spacetime and twistor 
space $Q$ one needs to substitute (\ref{twis}) into (\ref{omega}).

Consider the set of linear equations
\bea
\left[rD_r-2(\lm-u)D_u-\Phi\right]\Psi&=&0\nonumber\\
\left[2D_v+\fr{\lm-u}{r}D_r-\fr{\lm-u}{r^2}\Phi\right]\Psi&=&0.
\label{Lax}
\eea
Here $\lm \in \C$ and $(r,u,v)$ are the Poincar\'e coordinates which 
cover, only, the shaded region of FIG. \ref{fig-ads}.
The gauge fields $(\Phi,A_r,A_u,A_v)$ are $2\times 2$  
trace-free matrices depending only on $(r,u,v)$ and 
$\Psi(\lm,r,u,t)$ is a unimodular $2\times 2$ matrix function 
satisfying the reality condition $\Psi(\lm)\Psi (\bar{\lm})^\dagger
=I$ (where $\dagger$ denotes the complex conjugate transpose).
The system (\ref{Lax}) is overdetermined and in order for a solution
$\Psi$ to exist the following integrability conditions need to be satisfied
\bea
D_u \Phi&=&r F_{ur}\nonumber\\
D_v \Phi&=&-r F_{vr}\nonumber\\
D_r \Phi&=&-2r F_{uv}.
\label{sys}
\eea
The above equations are consistent with the ones obtained from (\ref{Bog}) 
using  the Poincar\'e coordinates.

The gauge and Higgs fields in terms of the function $\Psi$ can be obtained 
from the Lax pair (\ref{Lax}).
Note that, as  $\lm \ra \infty$ the function $\Psi$ goes to the identity 
matrix which implies that
\be
A_u=0,\hs\hs A_r=\fr{1}{r}\,\Phi.
\label{G1}
\ee
On the other hand, for $\lm=0$ and using (\ref{G1}) the rest of the 
gauge fields are defined as
\bea
\Phi&=&-\fr{r}{2}\,J_rJ^{-1}-u\,J_uJ^{-1}\nonumber\\
A_v&=&\fr{u}{2r}\,J_rJ^{-1}-J_vJ^{-1}
\label{G2}
\eea
where $J(r,u,v) \doteq \Psi(\lm=0,r,u,v)$.
Note that, in this case, the first equation of the system (\ref{sys}) is 
automatically satisfied (due to the specific gauge choice).

Recently, Ward \cite{W1} has shown that holomorphic vector
 bundles $V$ over $Q$
determine multi-soliton solutions of (\ref{sys}) in anti-de Sitter space
via the usual Penrose transform.
This way a five-parameter family of soliton solutions can be obtained, 
in a similar way as for flat spacetime \cite{W2}.
Later, more solutions of equations (\ref{sys}) were obtained 
by Zhou \cite{Zhou1,Zhou2} using Darboux transformations with constant
 and variable spectral parameters.
In what follows, we use the Riemann problem with zeros to 
construct families of soliton solutions and observe the 
occurence of different types of scattering behaviour.
More precisely, we present families of multi-soliton solutions
with trivial and nontrivial scattering.\\

{\large{\bf II. Construction of Solitons}}\\

The integrable nature of (\ref{Bog}) means that there is a variety of 
methods for constructing solutions.
Here, we indicate a general method for constructing soliton solutions
of (\ref{Bog}) which is a variant of that in Ref. \cite{W2}. 
Using the standard method of Riemann problem with zeros in order to 
construct the multi-soliton solution, we assume 
that the function $\Psi$ has the simple form in $\lm$, ie
\be
\Psi=I+\sum_{k=1}^n \fr{M_k}{\lm-\mu_k}
\label{Psi}
\ee
where $M_k$ are $2\times2$ matrices independent of $\lm$ and $n$ is the
soliton number.
The components of the matrix $M_k$ are given in terms of a rational function
$f_k(\omega_k)=a_k \,\omega_k+c_k$ of the complex variable:  
$\omega_k=v-r^2\,(\mu_k-u)^{-1}$.
Here $a_k$, $c_k$ and  $\mu_k$ are complex constants which determine 
the size, position and velocity  of the $k$-th solitons.
{\it Remark}: The rational dependence of the solutions $\Psi$ follows 
(directly) when the inverse spectral theory is considered.
In  \cite{FI} (for the flat spacetime), it was shown by solving the
Cauchy problem  that the spectral data is a function of a parameter
similar to  (\ref{omega}).

The matrix $M_k$ has the form
\be
M_k=\sum_{l=1}^n (\Gamma^{-1})^{kl} \bar{m}_a^l m_b^k
\ee
with $\Gamma^{-1}$ the inverse of 
\be
\Gamma^{kl}=\sum_{a=1}^2 (\bar{\mu}_k-\mu_l)^{-1}\bar{m}_a^km_a^l
\ee
and $m^k_a$ holomorphic functions of $\omega_k$, of the form
$m^k_a=(m_1^k,m_2^k)=(1,f_k)$.
The Yang-Mills-Higgs fields $(\Phi,A_r,A_v,A_u)$ can then be read off
from (\ref{G1}-\ref{G2}) and they automatically satisfy (\ref{sys}).
The corresponding solitons are spatially localized since $\Phi \ra 0$ 
 at spatial infinity (ie at $r=0$).

\begin{figure}
\begin{center}
\hskip 2.25cm
\put(98,150){$t=8$} 
\hskip 2.25cm
\epsfxsize=10cm\epsfysize=5cm\epsffile{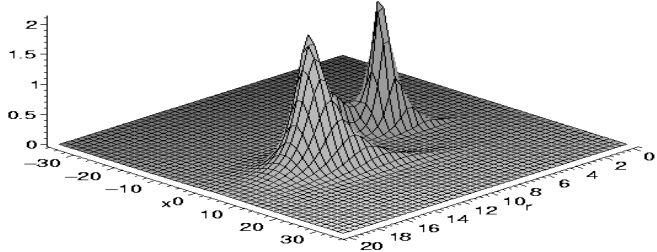}
\vskip -1cm
\end{center}
\caption{A two-soliton configuration at time $t=8$.}
\label{fig-triv}
\end{figure}  

By way of example, let us look at the special case where $\mu_1 =i$, 
$\mu_2=2i$, $a_1=2$, $a_2=1$, $c_1=5$ and $c_2=-10$.
FIG. \ref{fig-triv} represents a snapshot of the positive definite
gauge quantity ($-\mbox{tr} \Phi^2$) at time  $t=8$.
The corresponding solution consists of two solitons which travel towards 
$r=0$ and bounce back while their sizes change as they move.\\

{\large \bf III. Scattering Solutions}\\

The Riemann problem with zeros approach assumes that  the parameters $\mu_k$
 are distinct and also $\bar{\mu}_k \ne \mu_l$ for all $(k,l)$.
However, examples of generalizations of these constructions can be obtained
either involving higher order poles in $\mu_k$ or when $\bar{\mu}_k=\mu_l$.
When this procedure has been applied in  flat spacetime
 the corresponding solitons scatter in a nontrivial way. 
In particular, as it has been shown in \cite{W3,Ioan}, in head-on collisions
of $N$ indistinguishable solitons the scattering angle of the emerging solitons
relative to the incoming ones is $\pi/N$.
As a result, it would be of great interest to see the scattering behaviour
of the corresponding solitons in the anti-de Sitter spacetime.
Note that,  it is not clear  what to expect as a nontrivial scattering
(for example, $90^0$ scattering) in the shaded region of FIG. \ref{fig-ads}.
Another interesting point to be considered is the extension of the 
corresponding solutions to the whole anti-de Sitter space, ie the plot of
 the corresponding configuration in terms of the coordinates
 $(\rho,\theta,\phi)$.
This issue  will be addressed towards the end of the paper.

$\bigcirc$\, Firstly, let us look at an example in which the function $\Psi$
 has a double  pole in $\lm$ and no others.
In this case,  $\Psi$ has the form
\be
\Psi=I+\sum_{k=1}^2 \fr{R_k}{(\lm-\mu)^k}
\ee
where $R_k$ are $2\times 2$ matrices independent of $\lm$.
Then, as in flat spacetime \cite{W3}, $\Psi$ corresponds to 
a solution of (\ref{Lax}) if and only if it factorizes as
\be
\Psi(\lm)\!=\!\left(1-\fr{\bar{\mu}-\mu}{(\lm-\mu)} \fr{q^\dagger
 \otimes q}{|q|^2}\right) \!\left(1\!-\!\fr{\bar{\mu}-\mu}{(\lm-\mu)}
\fr{p^\dagger \otimes p}{|p|^2}\right)
\label{Psi-sca}
\ee
for some two vectors $q$ and $p$. One way to obtain the form of these 
vectors is by taking the formula (\ref{Psi}) for $n=2$ and setting
$\mu_1=\mu+\epsilon$, $\mu_2=\mu-\epsilon$, $f_1(\omega_1)=
f(\omega_1)+\epsilon h(\omega_1)$, $f_2(\omega_2)=f(\omega_2)-\epsilon 
h(\omega_2)$,  with $f$ and $h$ being rational function of one variable. 
In the limit $\epsilon \ra 0$ the two vectors $q$ and $p$
can be obtained  and are of the form:
\bea
q&=&(1+|f|^2)(1,f)+(\bar{\mu}-\mu)\! \left(\fr{r^2\,f'}
{(\mu-u)^2}+h\!\right)
(\bar{f},-1)\nonumber\\
p&=&(1,f).
\label{pq}
\eea
In this case, the constraint $f_2(\omega_2)-f_1(\omega_1)\ra 0$ 
as $\epsilon \ra 0$ has to be imposed in order for the resulting 
solution $\Psi$ to be smooth for  all $(r,u,v)$, which is 
true due to (\ref{omega}).
Note that the solution depends on the parameter $\mu$ and on the two
arbitrary functions $f$ and $h$.

Another way to obtain the aforementioned  solutions is by using the
Uhlenbeck construction  \cite{Uhl}; ie by assuming that the
 function $\Psi$ is a product of projectors  which 
satisfy first-order partial differential equations  and 
can easily be solved \cite{IZ}.

In order to illustrate the above family of solutions,  two
simple cases are going to be examined, by giving specific values to the
parameters $\mu$, $f(\omega)$ and $h(\omega)$.

\begin{figure}
\hskip 1.2cm
\put(100,90){$t=0$} 
\hskip 2.6cm
\epsfxsize=6.7cm\epsfysize=3cm\epsffile{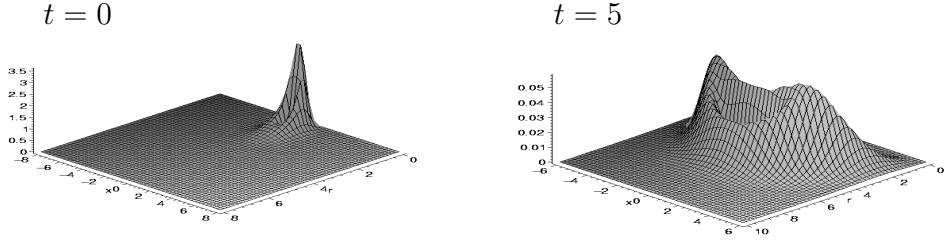}
\hfill
\hskip .1cm
\put(23,90){$t=5$}
\epsfxsize=6.7cm\epsfysize=3cm\epsffile{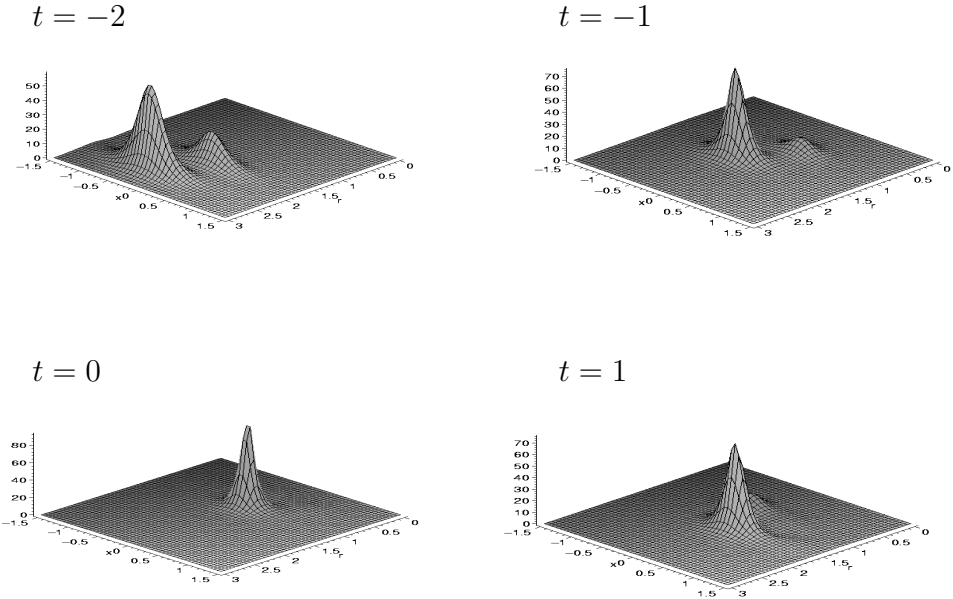}
\vskip -.5cm
\caption{A soliton configuration at different times.}
\label{fig-station}
\end{figure}  

(i) Let us study the simple case, where $\mu=i$, $f(\omega)=\omega$ and 
$h(\omega)=0$.
Then, the quantity $-\mbox{tr} \Phi^2$ simplifies to
\bea
-\mbox{tr}\Phi^2=32r^2 
\fr{[(r^2+x^2\!-\!t^2\!+\!1)^2+4t^2]
[(r^2+x^2\!-\!t^2\!-\!1)^2+4x^2]}{\left\{[(r^2+x^2-t^2)^2+1+2t^2+2x^2]^2+4r^4
\right\}^2},
\eea
which is time reversible.
The basic characteristics of the time-evolution of the above solution
are given qualitatively by FIG. \ref{fig-station}.
The time-dependent solution is a traveling soliton configuration which for
negative $t$, goes towards spatial infinity $(r=0)$; approaches it at 
$t=0$  and then bounces back at positive $t$.
During this period the soliton configuration deforms.

\begin{figure}
\vskip .25cm
\hskip .2cm
\put(100,90){$t=-2$} 
\hskip 2.6cm
\epsfxsize=6.7cm\epsfysize=3cm\epsffile{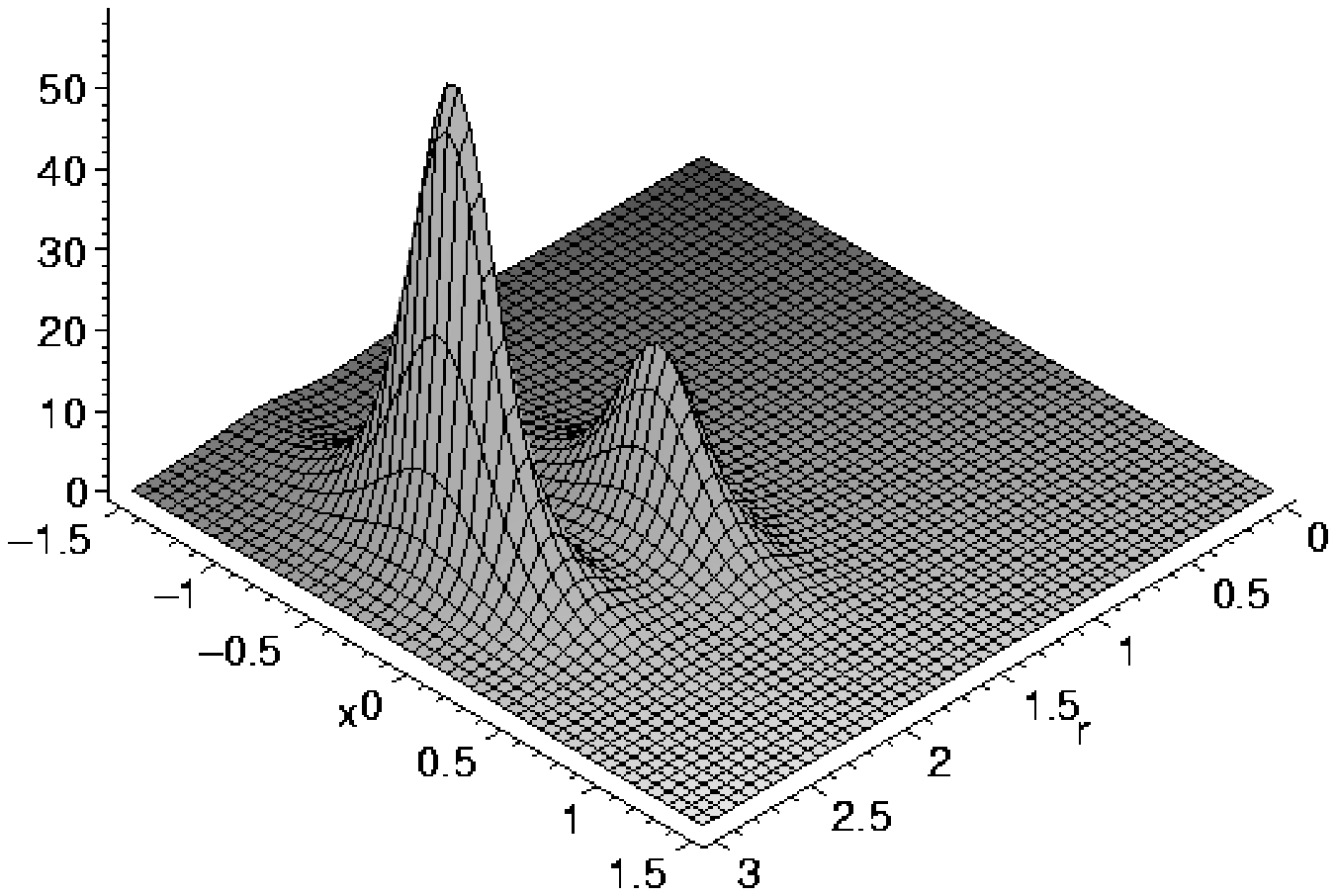}
\hfill 
\hskip 0.1cm
\put(29,90){$t=-1$}
\epsfxsize=6.7cm\epsfysize=3cm\epsffile{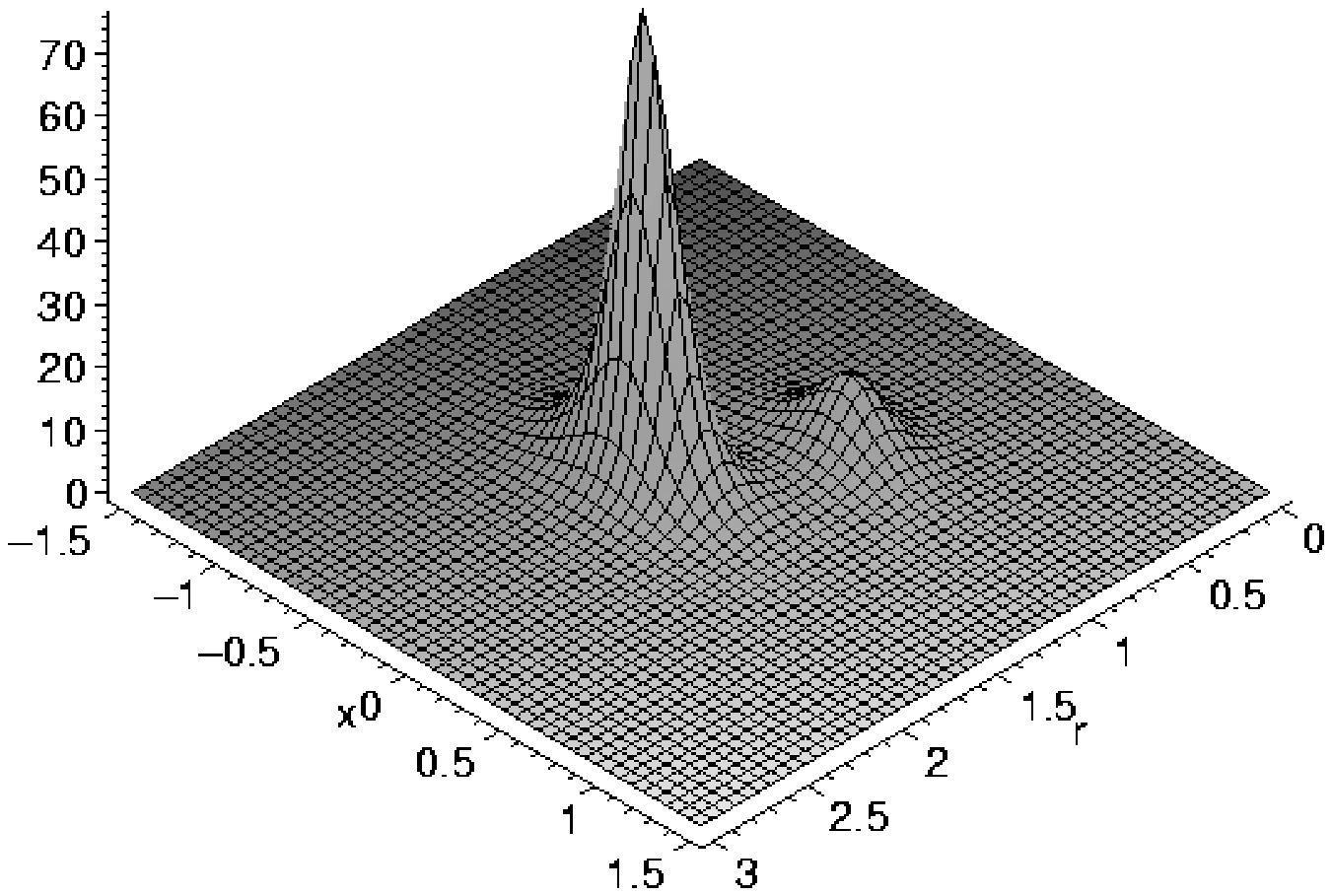}
\vskip -1cm
\vskip 2.25cm
\hskip 1.2cm
\put(100,90){$t=0$} 
\hskip 2.6cm
\epsfxsize=6.7cm\epsfysize=3cm\epsffile{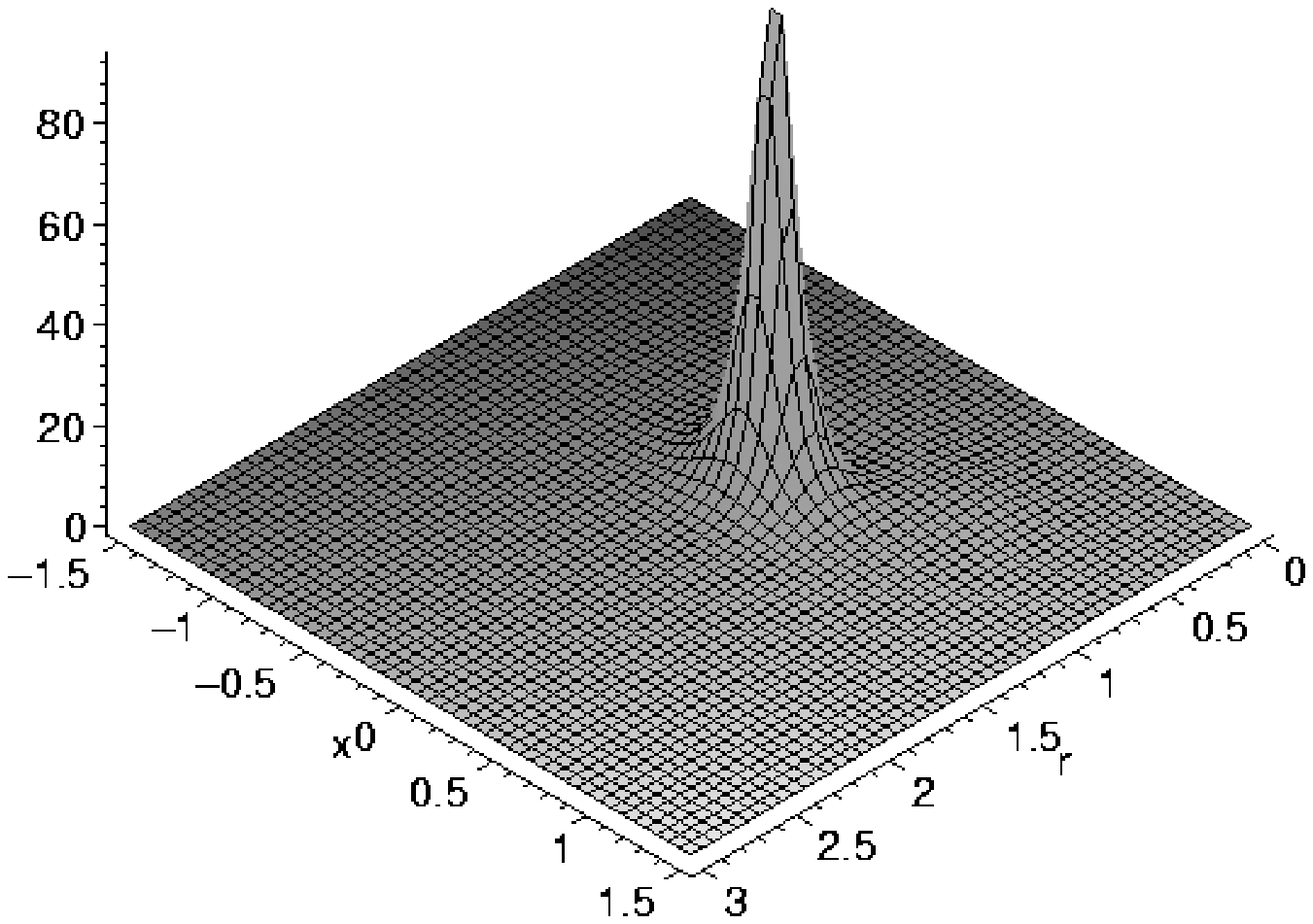}
\hfill 
\hskip 0.1cm
\put(29,90){$t=1$}
\epsfxsize=6.7cm\epsfysize=3cm\epsffile{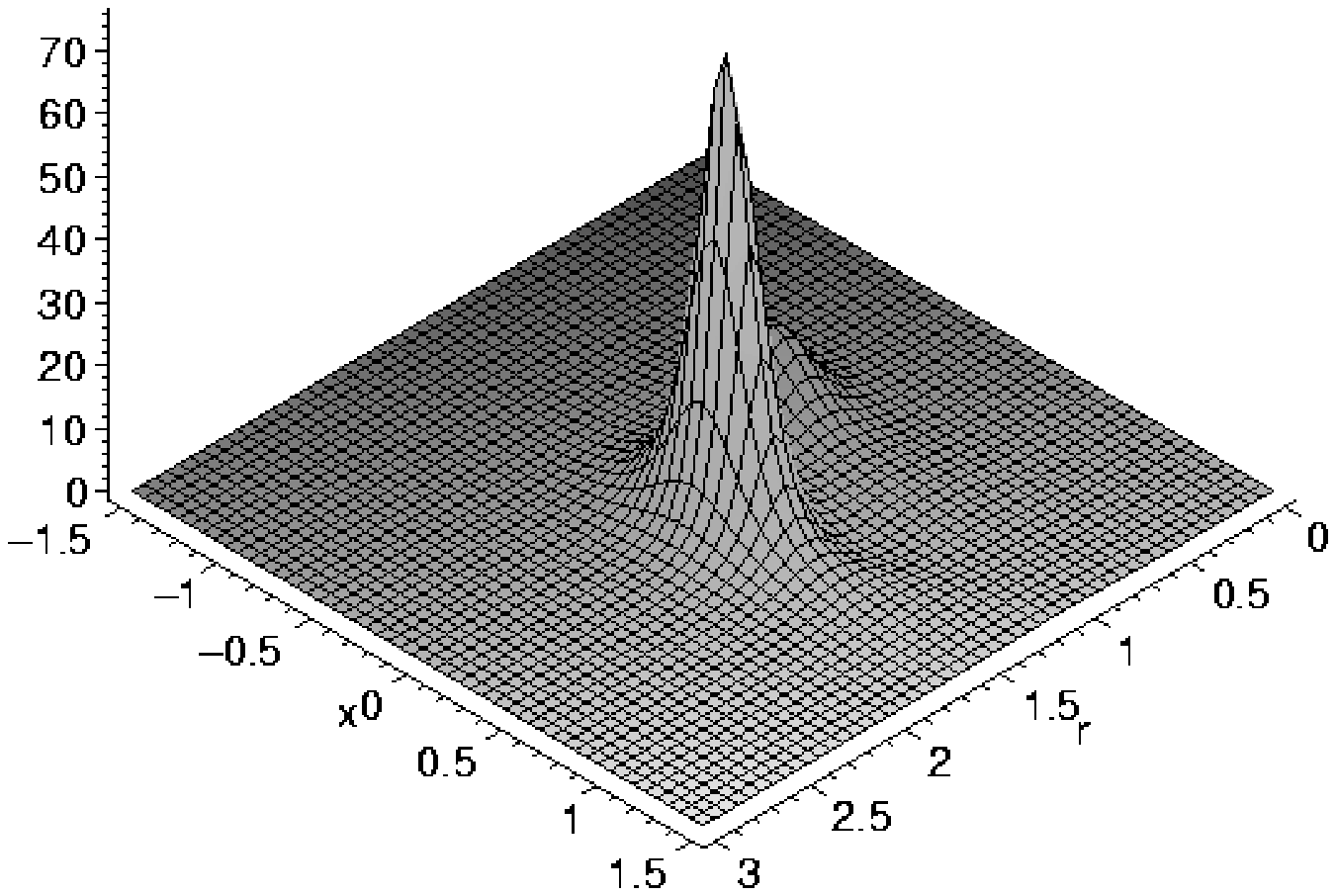}
\vskip -.5cm
\caption{A soliton configuration at different times.}
\label{fig-scat}
\end{figure}

(ii) Next, we investigate the solution which corresponds to a nontrivial 
scattering, at least in the flat spacetime.
FIG. \ref{fig-scat} represent the solution given by 
(\ref{Psi-sca}-\ref{pq}) for $\mu=i$, $f(\omega)=\omega$ 
and $h(\omega)=\omega^4$. 
The picture consists of two solitons with nontrivial scattering since,
for large (negative) $t$, the $-\mbox{tr} \Phi^2$ is peaked at two points 
which changes to a lump at $t=0$ and then two solitons emerge, for large
(positive) $t$, with the small one been shifted to the left.

This method can be extended to derive solutions which correspond
to the case where the function $\Psi$ has higher order pole in $\lm$
(and no others).
Then, $\Psi$ can be written as a product of three (or more) factors with three 
(or more) arbitrary vectors (for more details, see \cite{Ioan}).

$\bigcirc$\, Secondly, let us construct a large family of solutions which
correspond to the case where $\bar{\mu}_k=\mu_l$.
One way of proceeding is to take the solution $(\ref{Psi})$ with $n=2$,
put $\mu_1=\mu+\epsilon$, $\mu_2=\bar{\mu}-\epsilon$ and take the limit 
$\epsilon \ra 0$.
In order for the resulting $\Psi$ to be smooth it is necessary to take 
$f_1(\omega_1)=f(\omega_1)$, $f_2(\omega_2)=-1/f(\omega_2)-\epsilon h(\omega_2)$, where $f$ and
$h$ are rational functions of one variable.
On taking the limit we obtain a solution $\Psi$ of the form
\be
\Psi=I+\fr{n^1 \otimes m^1}{\lm-\mu}+\fr{n^2 \otimes m^2}{\lm-\bar{\mu}}
\label{sol-ant}
\ee
where $n^k$, $m^k$ for $k=1,2$ are complex valued two vector functions 
of the form
\bea
&&\hskip -1.2cm m^1=(1,f),\hs  m^2=(-\bar{f},1)\nonumber\acc
\hskip -0cm\pmatrix{\!n^1\cr n^2\!}\!\!&=&\!\!
\fr{2(\mu-\bar{\mu})}{4(1\!+\!|f|^2)^2\!-\!(\mu\!-\!\bar{\mu})^2|w|^2}
\!\!\pmatrix{\!2(1\!+\!|f|^2) \!&\! -(\mu\!-\!\bar{\mu}) \bar{w}\cr 
\!\!(\mu\!-\!\bar{\mu})w \!&\! -2(1\!+\!|f|^2)\!}
\pmatrix{\!m^{1 \dagger} \cr m^{2 \dagger}\!}
\eea
with 
\be
w\equiv \fr{2r^2}{(\mu-u)^2}f'+\bar{h}f^2.
\ee
So we generate a solution which depends on the parameter $\mu$ and the
two arbitrary rational functions $f=f(\omega)$ and $h=h(\bar{\omega})$.

\begin{figure}
\vskip 0.25cm
\hskip 1.2cm
\put(100,90){$t=-3$} 
\hskip 2.6cm
\epsfxsize=6.7cm\epsfysize=3cm\epsffile{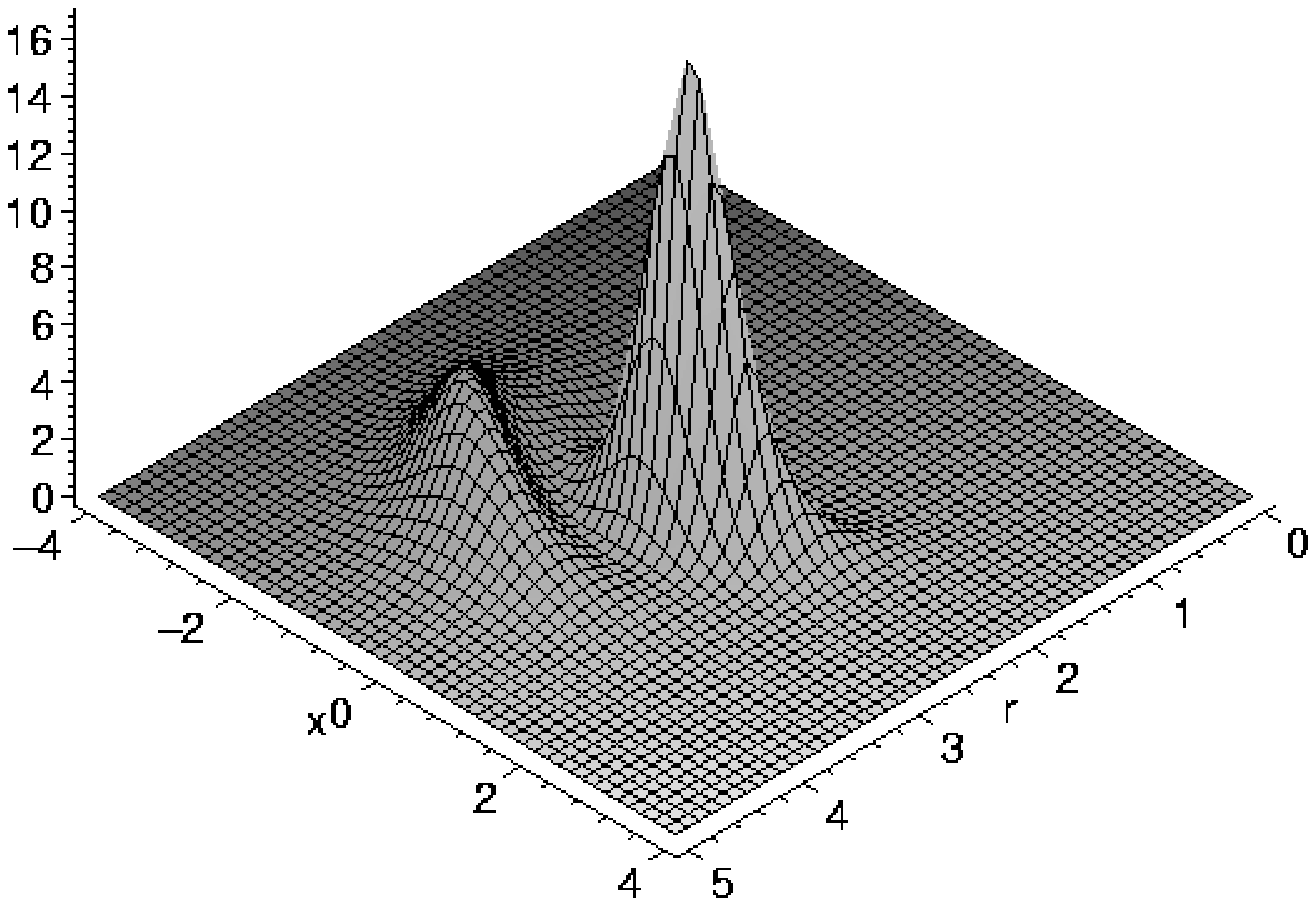}
\hfill 
\hskip .2cm
\put(29,90){$t=-1$}
\epsfxsize=6.7cm\epsfysize=3cm\epsffile{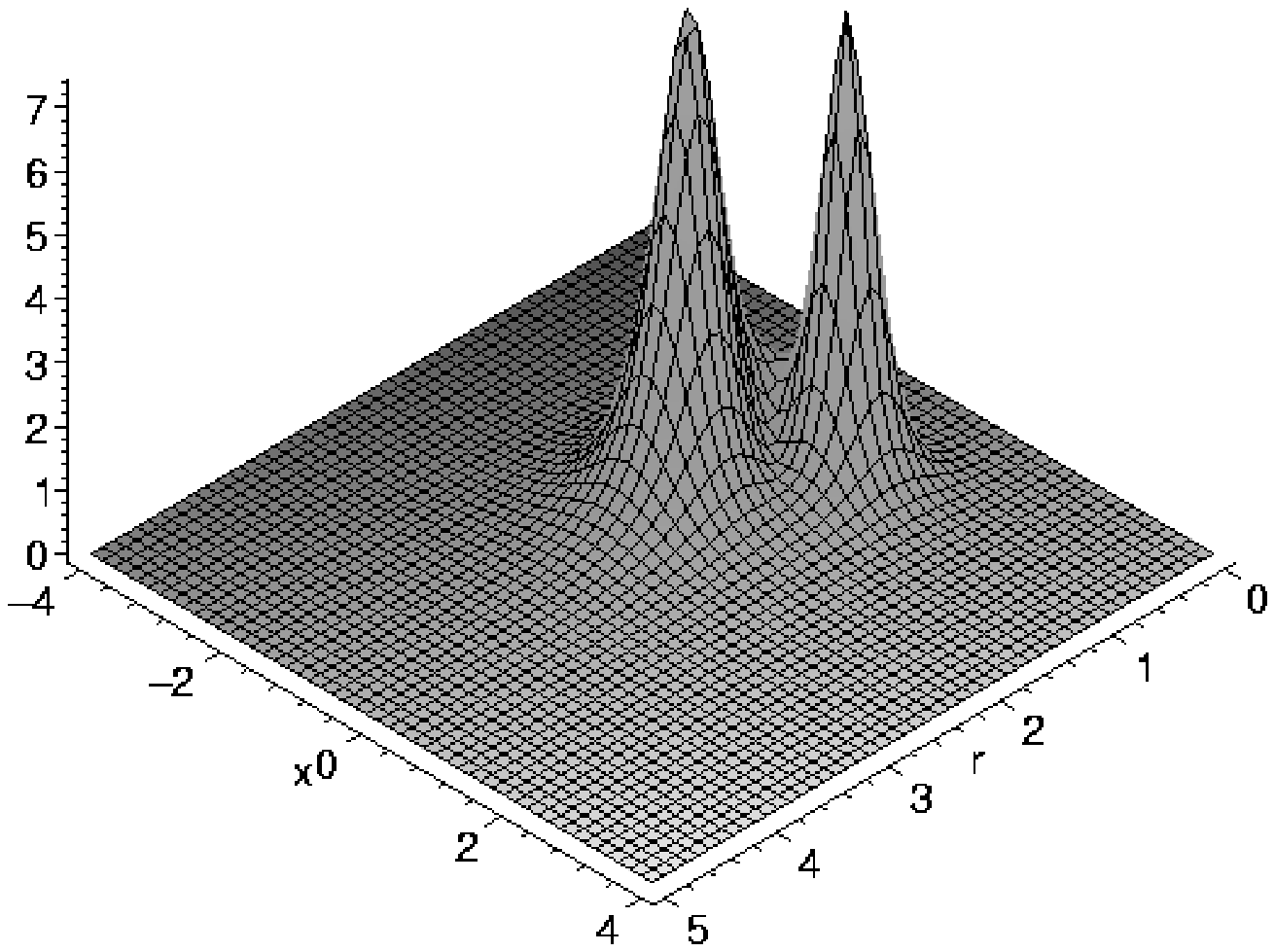}
\vskip -1cm
\vskip 1.25cm
\hskip 1.2cm
\put(100,90){$t=0$} 
\hskip 2.6cm
\epsfxsize=6.7cm\epsfysize=3cm\epsffile{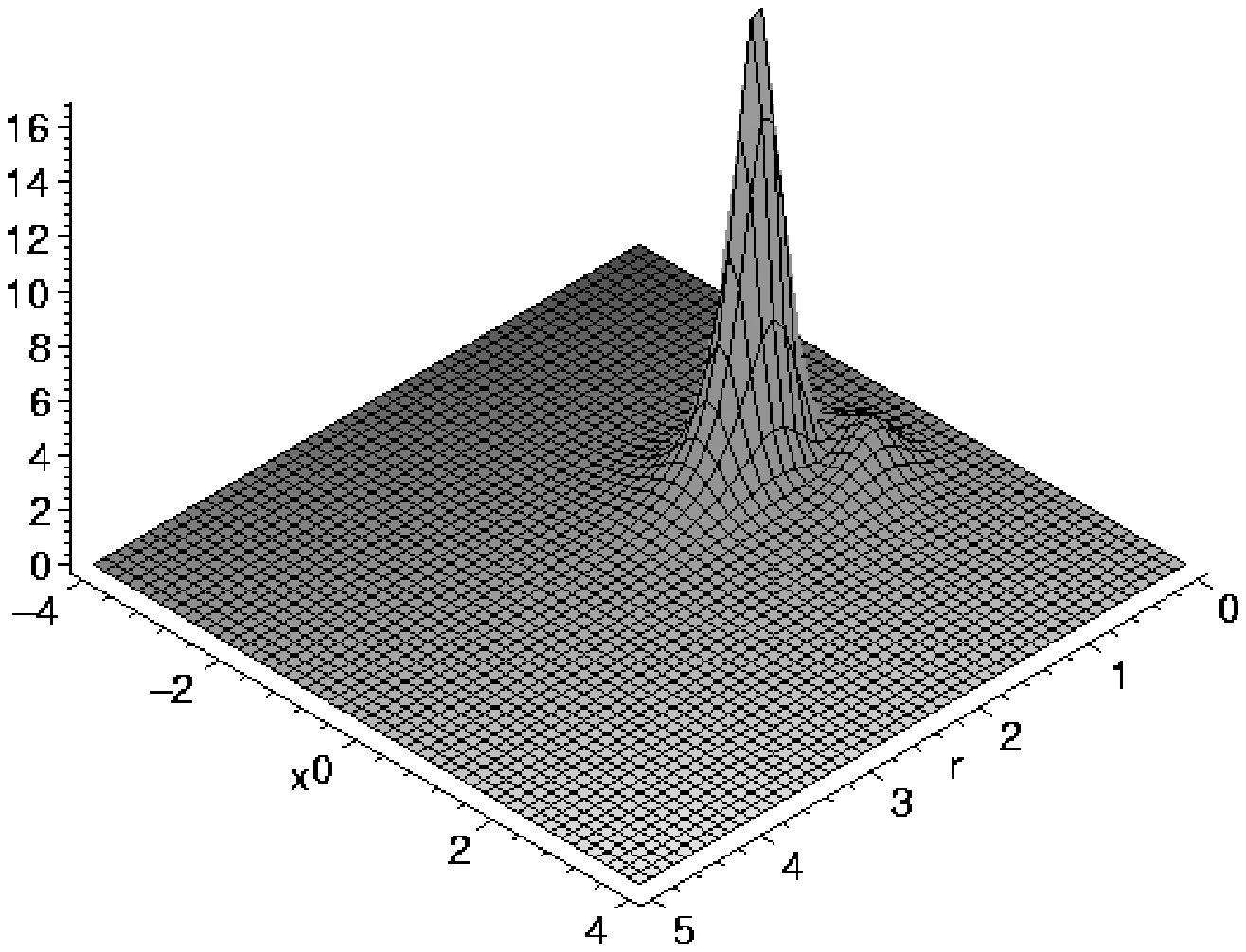}
\hfill 
\hskip .2cm
\put(29,90){$t=3$}
\epsfxsize=6.7cm\epsfysize=3cm\epsffile{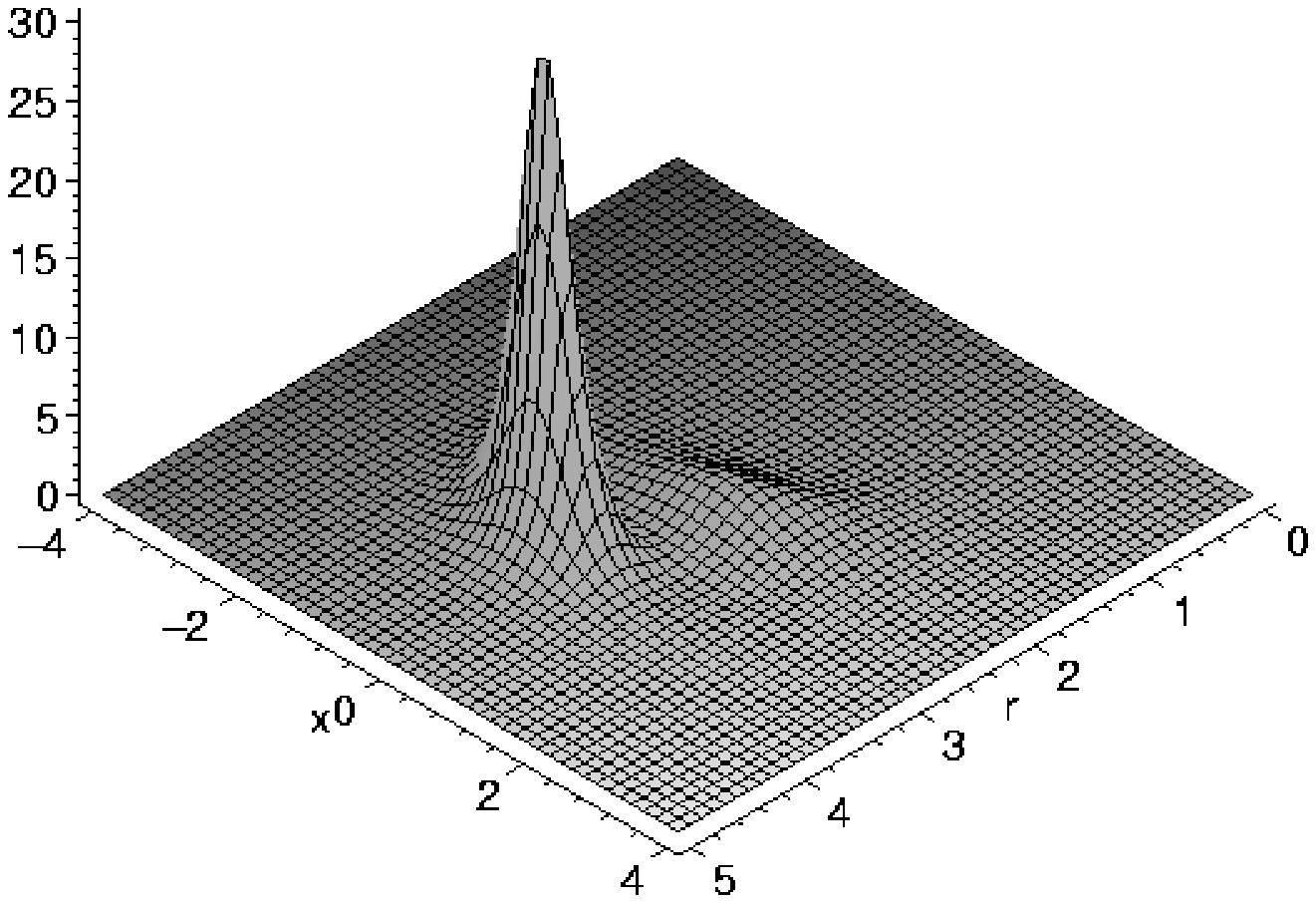}
\vskip -0.5cm
\caption{A soliton configuration at different times.}
\label{fig-sol-ant}
\end{figure}

In FIG. \ref{fig-sol-ant} we represent snapsots of the solution 
(\ref{sol-ant}) for $\mu=i$, $f=\omega$, $h=\bar{\omega}$.
The configuration consists of two solitons with nontrivial 
scattering behaviour.
Again, the quantity $-\mbox{tr} \Phi^2$ is peaked at two points,
for (negative) $t$, which are still distinct at $t=0$  and then two shifted 
(compared to the initial ones at $t=-3$)
solitons emerge, for (positive) $t$. 
Throughout the time-evolution their sizes change.

Note that, the scattering solutions belong to a large family
since $f$ and $h$ can be taken to be any rational  
functions of $\omega$.\\

{\large \bf IV. Conclusions}\\

Currently a great deal of attention has been focused on anti-de Sitter
spacetimes since they arise naturally in black holes and $p$-branes.
For the case of Yang-Mills theory with ${\cal N}=4$ supersymmetries 
and a large number of colours it has been conjectured that 
gauge strings are the same as the fundamental strings 
 but moving in a particular curved space: the product of 
five-dimensional anti-de Sitter space and a five sphere 
\cite{Mal}.
Then, using Poincar\'e coordinates the anti-de Sitter solutions 
play the role of classical  sources for the boundary field correlators, 
as shown in \cite{Wit}; while extensions of the corresponding statements 
can be applied to gravity theories, like the 
black holes which arise in anti-de Sitter backgrounds.

In this paper, we illustrate the construction of time-dependent
solutions related to hyperbolic monopoles.
In particular, families of solutions of the Bogomolny Yang-Mills-Higgs
equations in the (2+1)-dimensional anti-de Sitter space have been
constructed and their dynamics has been in studied in some detail.
As a result, it would be interesting to understand the role of higher poles 
in algebraic-geometry approach like twistor theory (for example,
the function $\Psi$ (\ref{Psi-sca}) correspond to $n=2$ bundles),
 and also to investigate the construction of the corresponding solutions 
and their dynamics  in de Sitter space.
Finally, it would be interesting to extend our construction 
in higher dimensional gauged theories and investigate the scattering
behaviour of the corresponding classical solutions and, also, consider
and study its noncomutative version (see, for exmaple, Ref. \cite{LP}).

\begin{figure}
\begin{center}
\hskip 2.25cm
\put(1,150){$\mbox{-tr}\Phi^2$} 
\put(100,30){$\rho$} 
\hskip 2.25cm
\epsfxsize=5cm\epsfysize=5cm\epsffile{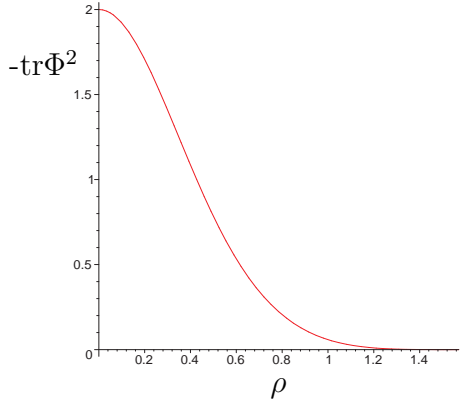}
\vskip -1cm
\end{center}
\caption{A one soliton configuration in the whole anti-de Sitter space.}
\label{fig-tel}
\end{figure}  

{\it Remark:}
The extension of  the  obtained classical solutions in the whole
anti-de Sitter space, ie using the coordinates $(\rho, \theta, \phi)$ 
is unambiguous.
For example, the simplest solution which corresponds to the one soliton 
(first derived in \cite{W1}) given by (\ref{Psi}) for $n=1$, 
$\mu_1=i$ and $f_1=\omega_1$ implies that 
\bea
-\mbox{tr}\,\Phi^2&=&\fr{8r^4}{[(r^2+x^2-t^2)^2+2x^2+2t^2+1]^2}\nonumber\\
&=&\fr{2 \cos^4\rho}{(\cos^2\rho-2)^2}
\eea
which means that  the positive definite quantity $-\mbox{tr}\,\Phi^2$
 is independent of the variables $(\theta,\phi)$ as shown 
in FIG. \ref{fig-tel}.

\section*{Acknowledgements}

Many thanks to Richard Ward for useful discussions and to the Nuffield
Foundation for a newly appointed lecturer award.
\\


\begin{thebibliography}{99}

\bibitem{At}
M. F. Atiyah,  Commun. Math. Phys. {\bf 93}, 437 (1984).

\bibitem{IS}
T. Ioannidou and P. M. Sutcliffe, J. Math. Phys. {\bf 40}, 5440 (1999).

\bibitem{JN}
S. Jarvis and P. Norbury, Bull. London Math. Soc. {\bf 29}, 737 (1997).

\bibitem{W1}
R. S. Ward, Asian J. Math. {\bf 3}, 325 (1999).

\bibitem{Gary}
G. W. Gibbons, {\it Anti-de Sitter Spacetime and its Uses}, Cambridge Preprint
(2001).
\bibitem{HE}
S. W. Hawking and G. F. R. Ellis, {\it The Large-Scale Structure of 
Space-Time}, CUP (1973).

\bibitem{Hitchin}
N. J. Hitchin, Complex manifolds and Einstein's equations. In: {\it Twistor
Geometry and Non-Linear Systems}, Eds H. D. Doebner and T. D. Palev (Lecture 
Notes in Mathematics 970, Springer-Verlag) (1982).

\bibitem{W2}
R. S. Ward, J. Math. Phys. {\bf 29}, 386 (1988).


\bibitem{Zhou1}
Z. Zhou, J. Math. Phys. {\bf 42}, 1085 (2001).

\bibitem{Zhou2}
Z. Zhou, J. Math. Phys. {\bf 42}, 4938 (2001).

\bibitem{FI}
A. S. Fokas and T. Ioannidou, Comm. Appl. Anal. {\bf 5}, 235 (2001).

\bibitem{W3}
R. S. Ward, Phys. Lett. A, {\bf 208}, 203 (1995).

\bibitem{Ioan}
T. Ioannidou, J. Math. Phys. {\bf 37}, 3422 (1996).

\bibitem{Uhl}
K. Uhlenbeck, J. Diff. Geom. {\bf 30}, 1 (1989).

\bibitem{IZ}
T. Ioannidou and W. J. Zakrzewski, J. Math. Phys. {\bf 39}, 2693 (1998).

\bibitem{Mal}
J. Maldacena, Adv. Theor. Math. Phys. {\bf 2}, 505 (1998).


\bibitem{Wit}
E. Witten, Adv. Theor. Math. Phys. {\bf 2}, 253 (1998).

\bibitem{LP}
O. Lechtenfeld and A. D. Popov, hep-th/0108118 (2001).


\end{thebibliography}
\end{document}